\documentclass[sigconf]{acmart}
\usepackage{balance}
\AtBeginDocument{%
  \providecommand\BibTeX{{%
    \normalfont B\kern-0.5em{\scshape i\kern-0.25em b}\kern-0.8em\TeX}}}

\copyrightyear{2024}
\acmYear{2024}
\setcopyright{acmlicensed}\acmConference[SIGIR '24]{Proceedings of the 47th International ACM SIGIR Conference on Research and Development in Information Retrieval}{July 14--18, 2024}{Washington, DC, USA}
\acmBooktitle{Proceedings of the 47th International ACM SIGIR Conference on Research and Development in Information Retrieval (SIGIR '24), July 14--18, 2024, Washington, DC, USA}
\acmDOI{10.1145/3626772.3661359}
\acmISBN{979-8-4007-0431-4/24/07}

\begin{document}

\title{A Preference-oriented Diversity Model Based on Mutual-information in Re-ranking for E-commerce Search}

\author{Huimu Wang}
\authornote{Both authors contributed equally to this research.}
\email{wanghuimu1@jd.com}
\author{Mingming Li}
\authornotemark[1]
\email{limingming65@jd.com}
\affiliation{%
  \institution{JD.com}
  \city{Beijing}
  \country{China}
}

\author{Dadong Miao}\authornote{Corresponding author.}
\email{miaodadong@jd.com}
\email{wangsonglin3@jd.com}
\email{tangguoyu@jd.com}
\author{Songlin Wang}
\author{Guoyu Tang}
\affiliation{%
  \institution{JD.com}
  \city{Beijing}
  \country{China}
}

\author{Lin Liu}
\email{liulin1@jd.com}
\email{xusulong@jd.com}
\email{hujinghe@jd.com}
\author{Sulong Xu}
\author{Jinghe Hu}
\affiliation{%
  \institution{JD.com}
  \city{Beijing}
  \country{China}
}







\renewcommand{\shortauthors}{Huimu Wang et al.}

\begin{abstract}
Re-ranking is a process of rearranging ranking list to more effectively meet user demands by accounting for the interrelationships between items. 
Existing methods predominantly enhance the precision of search results, often at the expense of diversity, leading to outcomes that may not fulfill the varied needs of users. Conversely, methods designed to promote diversity might compromise the precision of the results, failing to satisfy the users' requirements for accuracy.
To alleviate the above problems, this paper proposes a Preference-oriented Diversity Model Based on Mutual-information (PODM-MI), which consider both accuracy and diversity in the re-ranking process. Specifically, PODM-MI adopts Multidimensional Gaussian distributions based on variational inference to capture users' diversity preferences with uncertainty. Then we maximize the mutual information between the diversity preferences of the users and the candidate items using the maximum variational inference lower bound to enhance their correlations. Subsequently, we derive a utility matrix based on the correlations, enabling the adaptive ranking of items in line with user preferences and establishing a balance between the aforementioned objectives.
Experimental results on real-world online e-commerce systems demonstrate the significant improvements of PODM-MI, and we have successfully deployed PODM-MI on an e-commerce search platform. 
\end{abstract}






\maketitle

\section{Introduction}
Product Search Engine is a crucial component of E-commerce platforms.
To improve the efficiency of consumer decision-making during searches, a slate containing a limited number of ranked items is often presented in the search engine. 
In this case, most existing ranking methods \cite{cheng2016wide,guo2017deepfm,lian2018xdeepfm,zhou2019deep,gong2020edgerec,ai2018learning,zhuang2018globally,chen2021end,chen2022efficient,chen2022efficient,chen2018fast,carbonell1998use} often concentrate on predicting the click-through rate (CTR) for individual items, neglecting the critical mutual influence among contextual items and  the user's demands. As a result, a re-ranking stage is introduced to rearrange the initial list from the ranking stage to meet user demands. 
 
Nevertheless, User demands are inherently dynamic, shifting based on their search patterns and interactions. Searches for broad terms typically elicit a desire for a diverse range of options, while specific queries necessitate more targeted and precise results.
This behavior highlights the importance of considering both accuracy and diversity in re-ranking algorithms.

However, the existing methods often demonstrate divergent priorities. In terms of precision, certain methods tend to prioritize precision over other factors, such as RNN-based models \cite{bello2018seq2slate,zhuang2018globally,lin2022feature} . These methods suffer from feature information degradation with encoding distance and non-parallelizable characteristics caused by RNN \cite{rnn}. Conversely, methods like context-wise re-ranking strategy \cite{pei2019personalized,ai2018learning} leverage a context-wise evaluation model to capture the mutual influence among items and re-predict the CTR of each item. However, they do not directly model the diversity of user intent, resulting in a ranking that prioritizes precision over diversity and neglects the user's diverse needs. Regarding diversity, there exist effective diversity-promoting methods \cite{carbonell1998use,chen2018fast}. However, these methods rely on point-wise rank scores to measure the relevance between items and user interest, without considering the interplay between contents. Although methods such as DPP extension \cite{chen2018fast, gong2014practical, huang2021sliding}, have addressed this issue, balancing relevance and diversity still depends on hyper-parameter tuning and may not satisfy the diverse needs of different users. Moreover, methods based on greedy strategies may not obtain the global optimal solution.

To track these aforementioned issues, a novel Preference-oriented Diversity Model Based on Mutual-information (PODM-MI) is proposed in this paper. Specifically, it consists of two components including a preference-oriented network (PON) and a self-adaptive model(SAM). The PON is designed to model the representation of user diversity preference and the diversity within the candidate items. In contrast, the SAM dynamically adjusts the order of candidate items to align with the user's diversity preferences. 

Specifically, PON employs a multi-dimensional Gaussian distribution to model both the diversity preferences of users and the representation of diversity among candidate items, deviating from conventional embedding techniques, which yields two significant benefits: enhanced robustness compared to traditional embeddings and a more comprehensive characterization of diversity uncertainty with a higher-dimensional perspective. In the search scenario, history queries and the corresponding items of users are employed to represent the diversity preferences. Upon modeling the diversity representation of users and the candidate items, the SAM module adjusts the ranking of items to align with the user's diversity preferences. More precisely, SAM enhances the consistency between the diversity preferences of users and the diversity present in the candidate items. This consistency is accomplished by the maximization of mutual information between the user's diversity preferences and the candidate items' diversity. We introduce a variational posterior estimator to derive a tractable lower bound for the mutual information objective. This is also the first application of mutual information theory in the re-ranking model. After enhancing the consistency, a utility matrix predicated on the reinforced consistency is derived, which enables the adaptive re-ranking of items to reflect user preferences, thus maintaining equilibrium between diversity and accuracy. Moreover, the model is designed for straightforward implementation as an add-on module to any existing re-ranking algorithm.

The contributions of this paper can be summarized as follows:
\begin{itemize}
\item[$\bullet$] To the best of our knowledge, this is the first work that introduces mutual information to balance diversity and accuracy  for E-commerce re-ranking.
\item[$\bullet$] We propose a novel model that utilizes multi-dimensional Gaussian distribution to model diversity preference and adaptively matches the ranking results with user demands by maximizing the lower bound of mutual information.
\item[$\bullet$] We conduct extensive experiments on a real-world dataset.
Experimental results show that our model achieves significant
improvement over the state-of-the-art models. 
\item[$\bullet$] PODM-MI has been successfully deployed on the home search platform in JD and brought substantial economic benefits.

\end{itemize}

\section{PROBLEM FORMULATION}
Typically, The goal of the re-ranking is to elaborately select candidate items from the input ranking list and rearrange them into the final ranking list, followed by the exhibition for users. Mathematically, with a certain user $U$ and a ranking list of candidate items $C=\left \{ c_i \right \}_{i=1}^{N_C} $,  where $c_i$ is the $i-th$ item and $N_C$ is the total number of items in ranking list. 
The task of re-ranking stage is to learn a strategy $\pi^*$, $C\overset{\pi ^*}{\rightarrow} F$ which selects and rearranges items from $C$, and present a final ranking list $( F=\left \{ f_j \right \}_{j=1}^{N_F}, N_F < N_C )$ with the aim of meeting users demands such as diversity and accuracy.

\section{Methods}
\begin{figure}[t]
  \centering
  \includegraphics[scale=0.39]{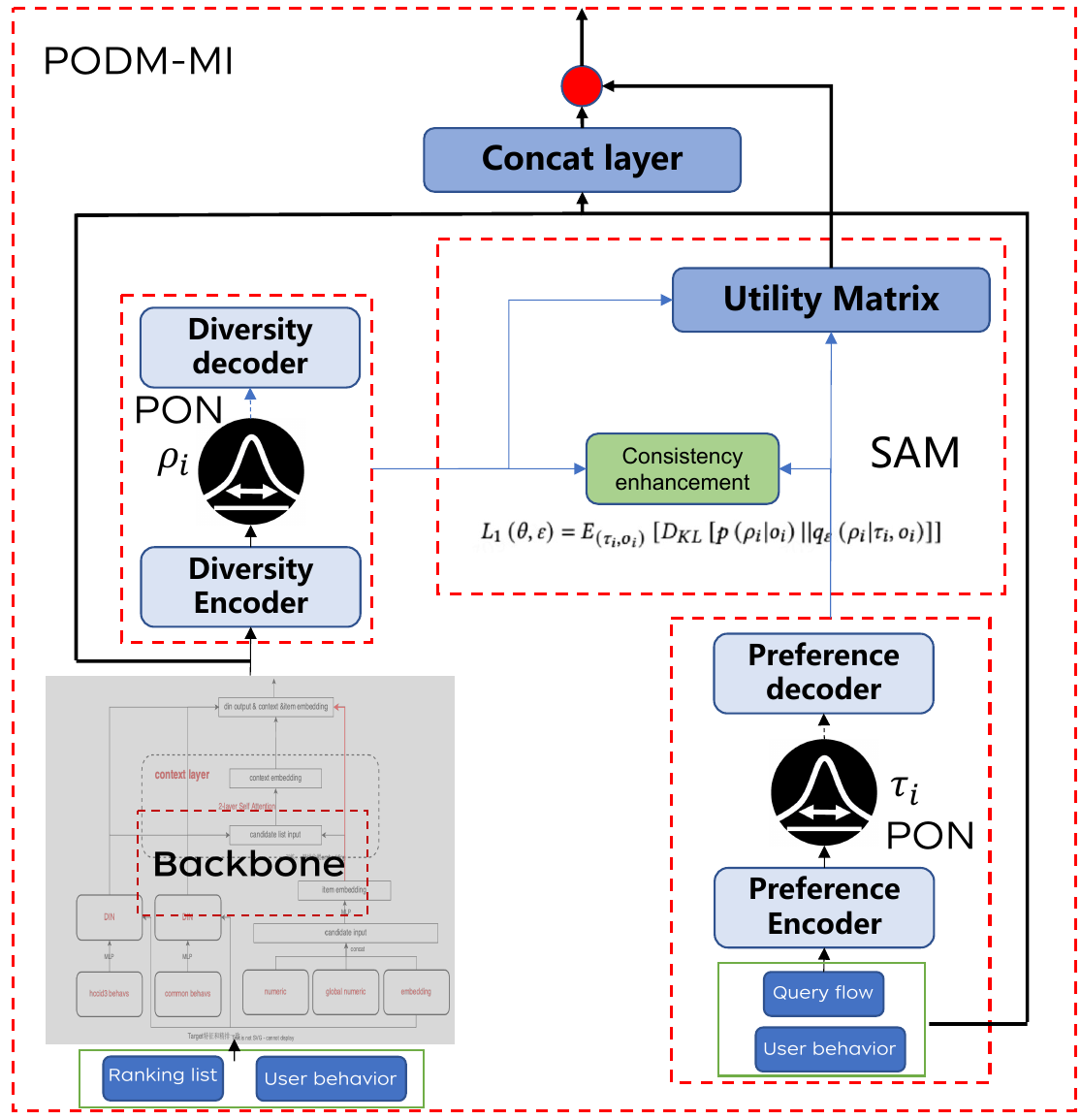}
  \caption{The framework of PODM-MI}
  \label{tod1}
  \vspace{-2.0em}
\end{figure}

\subsection{Overview}
We present the overview structure of PODM-MI consisting of preference-oriented network (PON) and a self-adaptive model(SAM) in Figure~\ref{tod1}. Specifically, we take ranking list $C=\left \{ c_i \right \}_{i=1}^{N_C} $ and user’s permutation-level behavior (such as click behavior flow, add-to-cart behavior flow) as the input of PODM-MI.
Then PON is utilized to model both the diversity preferences of users and the representation of diversity among candidate items. Next, SAM enhances the consistency between the diversity preferences of users and the diversity present in the candidate items. Subsequently, a utility matrix is then derived from this enhanced correlation, influencing the scores produced by the backbone network. This matrix is dynamically tuned to user preferences, facilitating the reordering of final ranking outcomes to better satisfy user preferences. The detailed
structure of PON and SAM will be introduced separately in the following sections.

\subsection{Preference-oriented Network}


In the e-commerce search scenario, historical queries and their associated items provide an effective representation of user intent. Thus, our methodology not only employ typical behavioral patterns (click behavior flow, add-to-cart behavior flow,etc.) but also includes query trajectories to enhance the depiction of user preferences.

After determining the features of the user preferences modeling, the next step is to determine the modeling method for user preferences representation. 
Traditional models tend to treat the dynamism of the user preferences as deterministic, yielding static user embeddings within a latent space. This approach is inadequate for capturing the complex nature of user preferences in dynamic real-world contexts.
Note that distribution representations introduce uncertainties and provide more flexibility compared with one single fixed embedding. Previous works [20, 29] have also demonstrated the superiority of representing user preferences as a distribution rather than an embedding. 

Therefore, we adopt a multi-dimensional Gaussian distribution, as delineated in Equation \ref{eq1}, to model the evolving trends of the user preferences.  The multi-dimensional Gaussian distribution governed by a mean vector and one diagonal covariance matrix. To be specific, user intent has two embedding representations, which are for mean ${\mu}$ and covariance ${\sum}$ that obtained from preference encoder. 
\begin{equation}
\label{eq1}
\resizebox{.9\hsize}{!}{
$\tau  \left ( x \right ) = \frac{1}{\left ( 2\pi  \right )^{\frac{2}{n} }\left | \sum_1  \right | ^{\frac{1}{2} } }exp\left ( -\frac{1}{2}\left ( x-\mu_1  \right )^{T}(\sum)^{-1}_1\left ( x-\mu_1  \right )  \right )$} 
\end{equation}
where ${x}$ is a n-dimensional vector representing the user preferences, ${\mu_1}$ is the mean vector , and ${\sum_1}$ is the covariance matrix.

We also model the candidate items with multi-dimensional Gaussian distribution. 

\begin{equation}
\label{eq2}
\resizebox{.9\hsize}{!}{
 $\rho \left ( y \right ) = \frac{1}{\left ( 2\pi  \right )^{\frac{2}{n} }\left | \sum_2  \right | ^{\frac{1}{2} } }exp\left ( -\frac{1}{2}\left ( y-\mu_2  \right )^{T}(\sum)^{-1}_2\left ( y-\mu_2  \right )  \right ) $ }
\end{equation}
where ${y}$ is a n-dimensional vector representing ranking list, ${\mu_2}$ is the mean vector , and ${\sum_2}$ is the covariance matrix.

\subsection{Self-adaptive Model}  
Upon modeling both the user preferences and the diversity inherent in the candidate items, the next step is to consider how to make the ranking results highly correlated with user intent. 
Mutual information, which gauges the shared information between two variables, can be utilized to quantify the correlation between candidate items and user preferences.

By maximizing mutual information between the aforementioned two (as shown in (\ref{equ3})), the distribution of the candidate items is highly consistent with the distribution of user intent.
\begin{equation}
\label{equ3}
max I\left({\rho}_{i};{\tau}_{i}| {o}_{i}\right)
\end{equation}
where ${o}_{i}$ represents the given ranking list.

However, estimating and maximizing mutual information is often intractable. Drawing inspiration from the literature of variational inference, we introduce a variational posterior estimator to derive a tractable lower bound for the mutual information objective:
\begin{equation}
I\left({\rho}_{i};{\tau}_{i}| {o}_{i}\right) \geq {E}_{\left({\rho}_{i},{\tau}_{i},{o}_{i}\right)}\left[\mathrm{\log}\left(\frac{{q}_{\varepsilon}\left({\rho}_{i}| {\tau}_{i},{o}_{i}\right)}{p\left({\rho}_{i}| {o}_{i}\right)}\right)\right] 
\end{equation}
where $q(x)$ is the marginal distribution of 
$q\left ( x|y \right ) $is the conditional distribution of $X$ given $Y$, and 
$p\left ( y|x \right ) $ is the conditional distribution of $Y$ given $X$.

The objective function for maximizing the lower bound of mutual information can be formulated as:
\begin{equation}
L_{1}\left ( \theta ,\varepsilon  \right )  = E_{\left ( \tau _i,o_i \right ) }\left [ D_{KL}\left [ p\left ( \rho _i|o_i \right )||q_{\varepsilon}\left ( \rho _i|\tau _i,o_i \right )   \right ]  \right ] 
\end{equation}

After enhancing their consistency, we design a learnable utility matrix to further align the final ranking results with user preferences. The learnable matrix is obtained by taking the dot product between a learnable weight matrix and the features that have been aligned. Then we multiply the utility matrix with scores computed from backbone network. 

By adjusting the values of the matrix, we can control the relative importance of different items and trends in the ranking process. For example, if certain items are more suitable for user preferences, we can adjust the values of the matrix to give them more weight in the ranking process. This approach allows the ranking results to be adaptively adjusted according to user intent.

\subsection{Model Training}
The proposed model is easily deployed as a follow-up module after any ranking algorithm by directly using the existing ranking feature vectors. Therefore, The loss of mutual information can be considered as an auxiliaries loss. For example, we use the order data as label and minimize the loss function for a backbone (such as PRM\cite{pei2019personalized}):
\begin{equation}
L_2=-\sum \sum z_{i}log(P(z_i|Y;\theta )) 
\end{equation}
where $z_{i}$ is the label (the label is positive if the item is purchased)

Then the total loss for the training process is :
\begin{equation}
L_{total}= L_1 + \chi L_2
\end{equation}
where $L_1$ is the loss function of mutual information and $\chi$ is the hyper-parameter.

\begin{table*}[th]
\caption{Overall performance comparison on the different methods in terms of $auc_{ord}@k$ and $entropy@k$.}
\label{exp1}
\begin{tabular}{cccccccccc}
\hline
Methods   & Auc\_ord@1      & Auc\_ord@3      & Auc\_ord@5      & Auc\_ord@10     & NDCG            & Brand@10        & Brand@20        & Shop@10         & Shop@20         \\ \hline
PRM       & 0.4236          & 0.6022          & 0.7176          & 0.8386          & 0.6854          & 1.2926          & 1.6299336       & 1.6389594       & 2.1226752       \\
PRM+TODMI & 0.4271          & 0.6052          & 0.7206          & 0.8412          & 0.6882          & 1.2953          & 1.63303047      & 1.64076226      & 2.12479788      \\
Boost     & \textbf{+0.83\%} & \textbf{+0.50\%} & \textbf{+0.42\%} & \textbf{+0.31\%} & \textbf{+0.40\%} & \textbf{+0.21\%} & \textbf{+0.19\%} & \textbf{+0.11\%} & \textbf{+0.10\%} \\ \hline
EGR       & 0.4279          & 0.6052          & 0.7192          & 0.8428          & 0.6875          & 1.2965          & 1.6332          & 1.6406          & 2.1248          \\
PODM-MI & 0.4326          & 0.609           & 0.7217          & 0.845           & 0.6905          & 1.2997          & 1.6364          & 1.6427          & 2.1267          \\
Boost     & \textbf{+1.10\%} & \textbf{+0.63\%} & \textbf{+0.35\%} & \textbf{+0.26\%} & \textbf{+0.44\%} & \textbf{+0.25\%} & \textbf{+0.20\%} & \textbf{+0.13\%} & \textbf{+0.09\%} \\ \hline
\end{tabular}
\end{table*}

\section{Experiments}



\subsection{Datasets and Metrics}
We collect search logs of the user for over two months from JD.com, one of China's largest B2C e-commerce websites, with the dataset size exceeding one billion and encompassing hundreds millions of items and users.

For accuracy estimation, we choose Area Under ROC (Auc) to evaluate offline performance. $Auc_{ord}@k$ ($K \in\{1, 3,5, 10\}$) denote Auc for the top-k items with respect to actual purchasing behaviors. For diversity evaluation, we use entropy to evaluate the richness of top-k items. $Brand / Shop@k$ ($K \in\{10,20\}$) denote the entropy related to brand and store of items. Moreover, we use UCVR (User Conversion Rate), UCTR (User Click Rate) and UV\_add\_cart (User Add-to-cart Rate) to evaluate online performance.

\subsection{Baselines}
In the industrial field, there are many classical methods for re-ranking. The most widely used work could be divided into several backbones: list-wise, evaluator-generator,etc.  Without loss of generality, we select PRM \cite{pei2019personalized}
(considering the personalized information) as baseline with the backbone of list-wise, and EGR \cite{egr} as the representive of the backbone of evaluator-generator. Our method is a general model, which could be adapted in
various versions of re-ranking framework. In this paper, we chose the PRM with PODM-MI and EGR with PODM-MI for a fair comparison.

 
\subsection{Implementation Details}
To ensure a fair comparison among different methods, we keep the feature columns, vocabulary size, the dimension of query/item unchanged. Specifically, we set the batch size as 1024. The learning rate is searched from $10^{-4}$ to $10^{-2}$. All models use Adagrad as the optimizer. The hyper-parameter in loss function is set to 1. 

\subsection{Offline Experimental Results}
The experimental results are shown in \ref{exp1}. From the results,
we can conclude that PODM-MI achieves a significant improvement with different backbones. Specifically:
1) Regarding precision metrics, the PODM-MI model significantly outperforms the baseline model in terms of AUC, which demonstrates that the ranking results of the model can effectively meet users' needs and illustrates its superiority and effectiveness. 2) Regarding diversity, the PODM-MI model shows varying degrees of improvement in store and brand diversity compared to other methods, indicating that the model effectively improves the diversity of ranking results. It is noteworthy that while improving diversity, precision has also been significantly improved, indicating that our model has effectively balanced precision and diversity. 3) Regarding compatibility, the PODM-MI model shows significant improvements in both performance and diversity compared to the original backbone when deployed on different backbones, indicating that our model is modular and can be easily deployed as a follow-up after any ranking algorithm.

\subsection{Online Experimental Results}
\begin{table}[t]
\vspace{-0.5em}
\caption{Online performance of A/B tests in JD. p-value is obtained by t-test.}
\label{exp2}
\setlength{\tabcolsep}{0.9mm}{
\begin{tabular}{cccccc}
\hline
\textbf{} & \textbf{UCVR} & \textbf{UCTR} & \textbf{UV\_add\_cart} & \textbf{Brand@10} & \textbf{Shop@10} \\ \hline
Gain      & +0.66\%       & +0.24\%       & +0.35\%                & +0.26\%           & +0.20\%          \\
p-value & 0             & 0.01          & 0.01                   & 0                 & 0                \\ \hline
\end{tabular}
}
\vspace{-2.0em}
\end{table}
To investigate the effectiveness of the PODM-MI, we conduct online A/B testing in the JD e-commerce search engine. The online performance is compared
against the previous EGR model online and is averaged over four consecutive weeks. 
As shown in Table.\ref{exp2}, the performance of PODM-MI increases by 0.66\% in UCVR (p-value=0), 0.26\% in terms
of diversity (p-value=0), and 0.24\% in UCTR (p-value=0.01). Small p-value means statistically significant, which demonstrates that PODM-MI not only enhances the likelihood of user purchases but also augments the diversity of items displayed in search results. Note that every 0.10\% increase in UCVR or UCTR brings great revenues for the company, hence the improvement achieved by the PODM-MI was significant. 
Consequently, PODM-MI has been fully integrated into the JD e-commerce search engine.

\subsection{Correlation Visualisation}
\begin{figure}[t]
  \centering
  \includegraphics[scale=0.20]{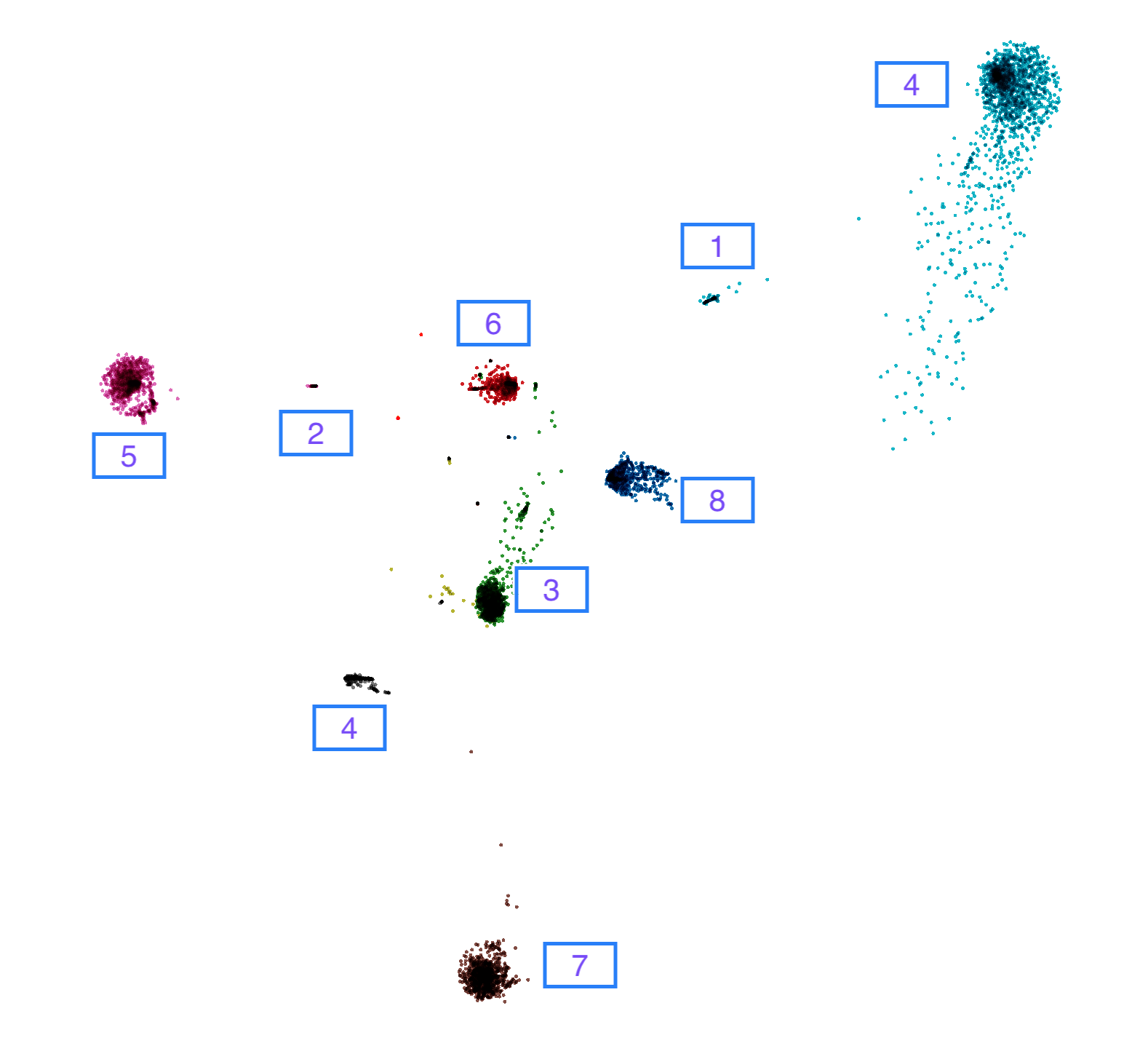}
  \caption{The correlation visualisation}
  \vspace{-2em}
\end{figure}
Distinct entropy values for the ranking results correspond to varied user intents. In order to assess whether the ranking results are high correlated with user intent, we visualized the distribution of entropy values of ranking results according to user intent using T-SNE \cite{van2008visualizing} dimensionality reduction. For enhanced clustering, we categorized the entropy levels into 8 distinct groups. 


The results presented in Figure 2 demonstrate that the clustering of user behavior flows with different levels of result diversity is highly noticeable, with well-defined boundaries for the distribution clusters of user intent. This indicates that the proposed model has successfully captured the underlying trends of user intentions and adjusted the ranking results accordingly. 

Notably, the entropy values of the ranking results increase as the user's intentions become more diverse, reflecting a higher degree of diversity in the ranking outcomes. Conversely, as the user's intentions become more precise, the entropy values decrease, indicating a higher level of accuracy in the ranking results.
~
~
~

\section{Conclusions}
This paper proposes a preference-oriented diversity model based on mutual-information, which harmonizes accuracy and diversity in re-ranking.  This is the first work that introduces mutual information to balance diversity and accuracy for E-commerce re-ranking. It consists of two components including PON and SAM. PON utilizes Multidimensional Gaussian distributions based on variational inference to model users’ diversity preferences with uncertainty.  SAM establishes the high correlation between the candidate items and user preferences by maximizing their mutual information.  Then utility matrix allows the ranking results to be adaptively adjusted according to user preferences. Extensive experiments conducted on a real-world dataset and the successful deployment on an e-commerce search platform show that the proposed framework can significantly surpass other state-of-the-art models.


\bibliographystyle{ACM-Reference-Format}
\balance
\bibliography{conference}


\appendix

\section{Company Portrait}
JD.com, Inc., also known as Jingdong, is a Chinese e-commerce company headquartered in Beijing. It is one of the two massive B2C online retailers in China by transaction volume and revenue, a member of the Fortune Global 500. When classified as a tech company, it is the largest in China by revenue and 7th in the world in 2021.

\section{Presenter Profiles}

\noindent\textbf{Huimu Wang} is a researcher in the Department of Search and Recommendation at JD.com Beijing. He received his doctor degree in Institute of Automation, Chinese Academy of Sciences. His research focuses on reinforcement learning and natural language processing.

\noindent\textbf{Mingming Li} is a researcher in the Department of Search and Recommendation at JD.com Beijing. He received his doctor degree in Institute of Information Engineering, Chinese Academy of Sciences. His research focuses on information retrieval.

\end{document}